%% file: main.tex
\documentclass{bmcart}

\usepackage[utf8]{inputenc} 
\usepackage[colorlinks,citecolor=blue,urlcolor=blue,bookmarks=false,hypertexnames=true]{hyperref} 

\usepackage{graphicx}

\startlocaldefs
\endlocaldefs

\begin{document}

\begin{frontmatter}

\begin{fmbox}
\dochead{Research}


\title{COVID-19 Tests Gone Rogue: Privacy, Efficacy, Mismanagement  and Misunderstandings}

\author[
   addressref={pcf, montreal},                   
]{\fnm{Manuel} \snm{Morales}}
\author[
   addressref={medialab, mit},                   
]{\fnm{Rachel} \snm{Barbar}}
\author[
   addressref={pcf},                   
]{\fnm{Darshan} \snm{Gandhi}}
\author[
   addressref={pcf},    
]{\fnm{Sanskruti} \snm{Landage}}
\author[
   addressref={pcf, stonybrook},              
]{\fnm{Joseph} \snm{Bae}}
\author[
   addressref={pcf},    
]{\fnm{Arpita} \snm{Vats}}
\author[
   addressref={pcf},                   
]{\fnm{Jil} \snm{Kothari}}
\author[
   addressref={pcf},                   
]{\fnm{Sheshank} \snm{Shankar}}
\author[
   addressref={pcf},                   
]{\fnm{Rohan} \snm{Sukumaran}}
\author[
   addressref={pcf},                   
]{\fnm{Himi} \snm{Mathur}}
\author[
   addressref={pcf},                   
]{\fnm{Krutika} \snm{Misra}}
\author[
   addressref={pcf,UCB},                   
]{\fnm{Aishwarya} \snm{Saxena}}
\author[
   addressref={pcf},                   
]{\fnm{Parth} \snm{Patwa}}
\author[
   addressref={pcf},                   
]{\fnm{Sethuraman} \snm{T. V.}}
\author[
   addressref={pcf, ITGH},                   
]{\fnm{Maurizio} \snm{Arseni}}
\author[
   addressref={pcf,NIH},                   
]{\fnm{Shailesh} \snm{Advani}}
\author[
   addressref={mit,pcf,harvardash},                   
]{\fnm{Kasia} \snm{Jakimowicz}}
\author[
   addressref={pcf},                   
]{\fnm{Sunaina} \snm{Anand}}
\author[
   addressref={pcf},                   
]{\fnm{Priyanshi} \snm{Katiyar}}
\author[
   addressref={pcf},                   
]{\fnm{Ashley} \snm{Mehra}}
\author[
   addressref={pcf},                   
]{\fnm{Rohan} \snm{Iyer}}
\author[
   addressref={pcf},                   
]{\fnm{Srinidhi} \snm{Murali}}
\author[
   addressref={pcf},                   
]{\fnm{Aryan} \snm{Mahindra}}
\author[
   addressref={pcf},                   
]{\fnm{Mikhail} \snm{Dmitrienko}}
\author[
   addressref={pcf},                   
]{\fnm{Saurish} \snm{Srivastava}}
\author[
   addressref={pcf},                   
]{\fnm{Ananya} \snm{Gangavarapu}}
\author[
   addressref={pcf},                   
]{\fnm{Steve} \snm{Penrod}}
\author[
   addressref={medialab, mit, harvardmed},                   
]{\fnm{Vivek} \snm{Sharma}}
\author[
   addressref={medialab, mit},                   
]{\fnm{Abhishek} \snm{Singh}}
\author[
   addressref={pcf, medialab, mit},                   
   corref={medialab},                       
   email={https://web.media.mit.edu/$\sim$raskar/}   
]{\fnm{Ramesh} \snm{Raskar}}


\address[id=pcf]{
  \orgname{PathCheck Foundation}, 
  \postcode{02139}                                
  \city{Cambridge},                              
  \cny{USA}                                    
}
\address[id=medialab]{%
  \orgname{MIT Media Lab},
  \postcode{02139}
  \city{Cambridge},
  \cny{USA}
}
\address[id=mit]{%
  \orgname{Massachusetts Institute of Technology},
  \postcode{02139}
  \city{Cambridge},
  \cny{USA}
}
\address[id=harvardmed]{%
  \orgname{Harvard Medical School},
  \postcode{02115}
  \city{Boston},
  \cny{USA}
}
\address[id=harvardash]{%
  \orgname{Ash Center for Democratic Governance and Innovation, Harvard Kennedy School},
  \postcode{02138}
  \city{Cambridge},
  \cny{USA}
}
\address[id=ITGH]{%
  \orgname{Institute for Technology and Global Health},
  \postcode{02139}
  \city{Cambridge},
  \cny{USA}
}
\address[id=NIH]{%
  \orgname{National Human Genome Research Institute, National Institutes of Health},
  \postcode{20892}
  \city{Bethesda},
  \cny{USA}
}
\address[id=UCB]{%
  \orgname{University of California, Berkeley School of Law},
  \postcode{94720}
  \city{Berkeley},
  \cny{USA}
}
\address[id=stonybrook]{%
  \orgname{Renaissance School of Medicine, Stony Brook University},
  \postcode{11794}
  \city{Stony Brook},
  \cny{USA}
}
\address[id=montreal]{
  \orgname{University of Montreal}, 
  \postcode{H3T 1J4}                                
  \city{Montreal},                              
  \cny{Canada}                                    
}


\begin{artnotes}
\note{\textsuperscript{1}PathCheck Foundation, 02139 Cambridge, USA.\\ \textsuperscript{2}MIT Media Lab, 02139 Cambridge, USA.\\ \textsuperscript{3}Massachusetts Institute of Technology, 02139 Cambridge, USA.\\ \textsuperscript{4}Harvard Medical School, 02115 Boston, USA.\\ \textsuperscript{5}Ash Center for Democratic Governance and Innovation, Harvard Kennedy School, 02138 Cambridge, USA.\\ \textsuperscript{6}Institute for Technology and Global Health, 02139 Cambridge, USA \\ \textsuperscript{7}National Human Genome Research Institute, National Institutes of Health, 20892 Bethesda, USA \\ \textsuperscript{8}University of California, Berkeley School of Law, 94720 Berkeley, USA\\ \textsuperscript{9}Renaissance School of Medicine, Stony Brook University, 11794 Stony Brook, USA\\ \textsuperscript{10}University of Montreal, H3T 1J4, 11794 Montreal, Canada} 
\end{artnotes}

\end{fmbox} 


\begin{abstractbox}

\begin{abstract} 
COVID-19 testing, the cornerstone for effective screening and identification of COVID-19 cases, remains paramount as an intervention tool to curb the spread of COVID-19 both at local and national levels. However, the speed at which the pandemic struck and the response was rolled out, the widespread impact on healthcare infrastructure, the lack of sufficient preparation within the public health system, and the complexity of the crisis led to utter confusion among test takers.  Invasion of privacy remains a crucial concern. The user experience of test takers remains low.  User friction affects the user behavior and discourages participation in testing programs.  Test efficacy has been overstated.  Test results are poorly understood resulting in inappropriate follow-up recommendations. Herein, we review the current landscape of COVID-19 testing, identify four key challenges, and discuss the consequences of the failure to address these challenges.

The current infrastructure around testing and information propagation is highly privacy invasive and does not leverage scalable digital components. In this work, we discuss challenges complicating the existing covid-19 testing ecosystem and highlight the need to improve the testing experience for the user and reduce privacy invasions. Digital tools will play a critical role in resolving these challenges.

\end{abstract}




\end{abstractbox}
%

\end{frontmatter}




\input{content/introduction}
\input{content/challenges_and_consequences}
\input{content/discussion}


\begin{backmatter}

\section*{Competing interests}
  The authors declare that they have no competing interests.

\section*{Acknowledgements}
  We are grateful to Riyanka Roy Choudhury, CodeX Fellow, Stanford University, Adam Berrey, CEO of PathCheck Foundation, Dr. Brooke Struck, Research Director at The Decision Lab, Canada, Vinay Gidwaney, Entrepreneur and Advisor, PathCheck Foundation, and Paola Heudebert, co-founder of Blockchain for Human Rights, Alison Tinker, Saswati Soumya, Sunny Manduva, Bhavya Pandey, and Aarathi Prasad for their assistance in discussions, support and guidance in writing of this paper.


\bibliography{refs}
\bibliographystyle{bmc-mathphys}


\end{backmatter}
\end{document}

%% file: content/introduction.tex
\section{Introduction}
Testing plays a consequential role in curbing the spread of COVID-19.  Testing can be carried out to identify cases and isolate the infected person so that the virus does not spread to others. Several testing methods have been developed and deployed, each with its own strengths and weaknesses. Data collection and generation are innate components of testing programs, but expose users to privacy invasions if not managed carefully.  Like all medical tests, COVID-19 tests are complicated by limitations on the sensitivity and specificity of test methods.  If not  adequately communicated, these limitations confuse test takers, healthcare providers, and public health officials.  Wait times, supply chain problems, and varied reporting requirements all further complicate a person’s ability to undergo testing and utilize the result of the test. In this draft, challenges in COVID-19 testing are grouped into four categories: Privacy, Efficacy, Workflow, and Communication.   In order to effectively curb the spread of COVID-19 while avoiding negative effects, it is then essential to understand further these notions and have inclusive discussions that would lead to designing more responsive, resilient, and societal sensitive systems. 

\subsection{Related Work}

High profile events have been held with people at the event later testing positive for COVID-19\cite{Tully}. As was the case with this high profile event, attendees underwent testing prior to the event and were then allowed to attend without a mask or other social distancing measures. Rapid antigen, antibody and RT-PCR tests vary in the probability for a false negative result for the SARS-CoV-2 virus.  Variation is impacted by the timing of the test with different false negative rates depending on whether the test was conducted before or after the appearance of symptoms\cite{Guglielmi}. The spread of COVID-19 infections among guests following the event highlights challenges with test efficacy. 

Universities and colleges bringing students back to campus this fall relied on COVID-19 tests to contain the spread of COVID-19 as the campus community resumed face to face learning.  Testing programs varied widely among universities, with some offering testing only to symptomatic individuals while others required testing prior to or after arrival on campus.  A few universities implemented surveillance testing on their campuses and made participation a requirement for attending class on campus. Results and efficacy of these programs have been as diverse as the design of the programs and provide good insight for communities, employers, and others considering implementation of a testing strategy. 

COVID-19 testing is not a standalone strategy, but rather one component amongst several to help mitigate spread.  Yet as several colleges and universities have shown, thoughtful, well-implemented testing programs have a significant role to play in the pandemic response.     

\subsection{Tests for COVID-19}
A number of tests are commercially available for COVID-19.  To date, reverse transcription polymerase chain reaction (RT-PCR) tests remain the “gold standard” offering the highest sensitivity and specificity. However, new antigen tests are dramatically revamping public health policy by offering results quickly. A detailed analysis of the clinical landscape of the tests has been completed by Gandhi et. al., \cite{gandhi2020clinical}

\begin{figure}[!ht]
    \includegraphics[width = .95\linewidth]{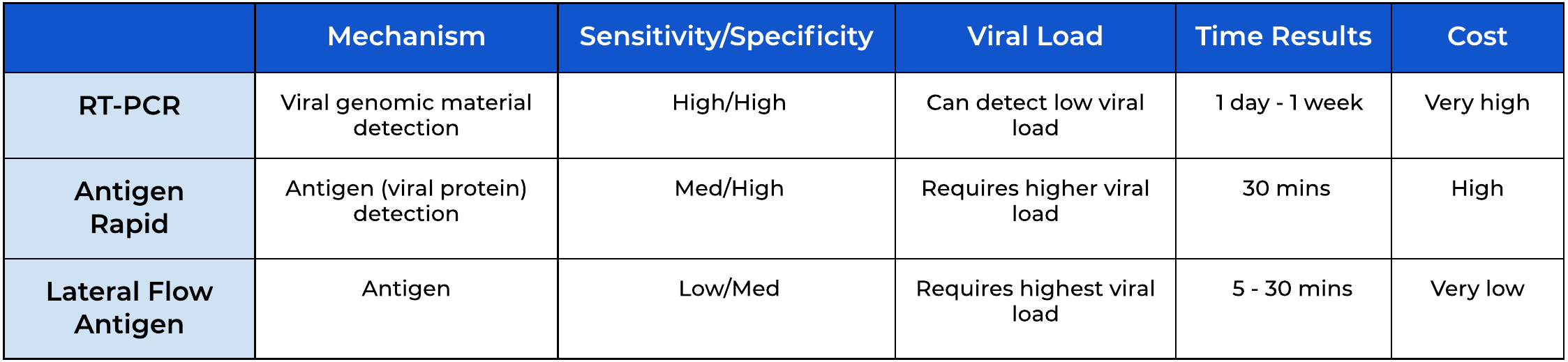}
    \centering
    \caption{Comparison of key traits across three primary testing methodologies \cite{gandhi2020clinical}}
    \label{fig:comparison_table}
\end{figure}

\textbf{RT-PCR}: Reverse transcription-polymerase chain reaction (RT-PCR) offers the highest sensitivity (percent of correctly classified true positives) and specificity (percent of correctly classified true negatives) of the currently available tests. RT-PCR identifies viral genomic material in samples taken from the nasopharyngeal cavity, throat, or mouth. \cite{Kumar} RT-PCR processing requires specialized laboratory-based equipment and trained technicians. Specimens must be shipped to centralized laboratories. While specimen processing and testing can be completed within a few hours of arrival at the laboratory, the total turnaround time to patients is typically 1-7 days.  \cite{Rahman} \cite{Mardani}

\textbf{Rapid Antigen Detection Tests}: Rapid Antigen Detection Tests (RADT) detect the presence of virus-specific proteins (antigens) in specimens collected from the same sites as RT-PCR (nasopharyngeal, nasal, throat, etc.) using immunological assays in a point-of-care setting.  RADT requires commonly available laboratory equipment and reagents and materials are typically packaged in a single kit.  Results for RADT can be ready in 30 to 60 minutes.  The sensitivity of these tests is generally lower than RT-PCR, typically 84-97\% of RT-PCR results. However, these tests are considered very specific,  90-98 \cite{CDC1} 

\textbf{Lateral Flow Assay Antigen Tests}: Lateral flow assay (LFA) antigen tests are another example of rapid point-of-care testing. These assays analyze plasma, blood, or other patient samples for viral antigens and can be performed in any clinical setting. LFA antigen tests do not require laboratory-based equipment and obtain results faster than RADT. \cite{Green}

\textbf{At-Home Tests}: Several companies are working to develop rapid, at-home tests for COVID-19.  Such tests will expand access and reduce barriers to testing.  Questions about sensitivity and specificity have complicated the roll out of at-home tests.   Yet, at-home tests have received support from a bipartisan group of seven governors in the United States and are likely to increase in prevalence and importance in COVID-19 testing programs in the coming months.  \cite{Duncan} \cite{Baraniuk} \cite{McCullough}

\textbf{Antibody Tests and Community-Level Tests}: Antibody tests determine whether a patient has developed immunity to COVID-19 following a known or presumed infection. Antibodies are generated by the body as a physiological immune response to SARS-CoV-2 infection, and antibody tests measure the presence of IgM or IgG antibodies against the virus. \cite{FDA1}  The information provided by antibody testing is highly variable from patient to patient and depends significantly on when the test was acquired relative to infection with SARS-CoV-2.  Antibodies appear in patient samples 1 to 3 weeks following infection. Some jurisdictions use antibody testing to identify plasma donors for COVID-19 treatment.  While not the focus of this paper, antibody testing is likely to increase as vaccines become available and immunity to COVID-19 becomes more widespread. \cite{Kumar} \cite{Billingsley}

Waste-water testing, testing sewer water for the presence of the SARS-CoV.2 virus, as well as pool testing have been used to monitor the level of infection in the community rather than to identify infected individuals.  These tests have significant public health utility, but are beyond the scope of this current draft.  \cite{Watkins} \cite{FDA2} \cite{FDA3}

%% file: content/challenges_and_consequences.tex
\section{Challenges and Consequences}  
Testing technologies are rapidly evolving and so is our understanding of the contagion phenomenon and its timeline. Combined, these breakthroughs offer a powerful tool to mitigate the risk associated with the spread of the disease. New comprehensive strategies can now be designed to improve our current capabilities to fight the spread, however this is not a straightforward exercise. Challenges in deploying mitigation strategies in combination with test programs using new technologies for testing are numerous and they represent non-negligeable bottlenecks for the rolling out of such programs. Although tackling these bottlenecks require quick and decisive actions given the emergency situation, one needs to reflect as well at the unintended consequences of a strategy rollout that is not adequately addressing these challenges. In this section we review in some depth these challenges and we elaborate on the consequences of inadequately addressing them.  These challenges fall into four main categories: Privacy, Test Efficacy, Workflow, and Communication.
\begin{itemize}
    \item \textbf{Privacy}: Testing programs must collect private information to enable an adequate follow up. Mishandling of this data, an inadequate collection strategy or weak anonymity preservation protocols are some of the challenges around privacy.
    \item \textbf{Test Efficacy}: A comprehensive quantitative understanding of the thresholds and timelines for accuracy of tests as well as their false positives and false negatives rates must be integrated into a testing program roll out for it to be successful.
    \item \textbf{Workflow}: At the back-end of a testing program lies a supply chain workflow that requires efficient planning, scheduling and operationalization for it to have a real impact.
    \item \textbf{Communication}: How the overall objectives of a testing program within a mitigation strategy are presented is also crucial to its success. There are challenges around how to communicate to the public facts about test efficacy, interpretation of results, private data use, collection and handling as well as distribution of test, all within a unified and coherent message.
\end{itemize}

Incomplete or inadequate responses to these challenges have identifiable consequences that hinder the overall aimed objectives sometimes at great cost for the success of the strategy. Most crucial consequences fit into 6 main forms: Spread of Disease, Individual Behavior, Societal Impact, Economic Impact, Security and Bad actors and Technology Rollout.  
\begin{itemize}
    \item \textbf{Disease Spread}: The main objective of a testing program is to curb the spread. Partially addressed challenges will directly impact the ability of the strategy to mitigate the spread.
    \item \textbf{Individual Behavior}: A second objective of a mitigation strategy is to influence individual decision-making processes in their daily lives in a way that effectively gears collective behavior towards safety choices. Failure to address the above mentioned challenges will hinder the ability of the strategy to positively influence individual behavior hence multiplying spread events. 
    \item \textbf{Societal Impact}: Today’s societies exhibit systemic inequalities that have marginalized and disenfranchised communities and that represent by themselves challenges towards our pursuit towards social justice. Partially addressed challenges may create or exacerbate sigma and discrimination towards already marginalized segments of the population rendering the fabric of our democracies even more fragile. 
    \item \textbf{Economic Impact}: The managing of the current sanitary crisis has already come at great cost to our economies. Sacrifices have already been made in order to curb the spread and safe lives and hard choices still remain ahead of us. Adequately addressing these challenges might reduce the negative impact on our economies as testing programs are rolled out and integrated into mitigating strategies allowing for a safe path towards economic recovery. 
    \item \textbf{Security and Bad Actors}: Partially addressed challenges may open the door to abuse or misuse of the system and processes put in place. These can represent back-door opportunities for ill-motivated individuals or entities to misuse and exploit weaknesses and/or information for their own advantage or interest.
    \item \textbf{Technology Rollout}: Technology must play an important role as a vehicle to effectively operationalize a testing program and its successful integration into an overarching mitigation strategy. 
\end{itemize}

Challenges and the consequences of unaddressed challenges impact the user’s perception and experience of COVID-19 testing.  If the dynamic interrelationship between the challenges and consequences is overlooked, the testing program is compromised.

\begin{figure}[!ht]
    \includegraphics[width = .95\linewidth]{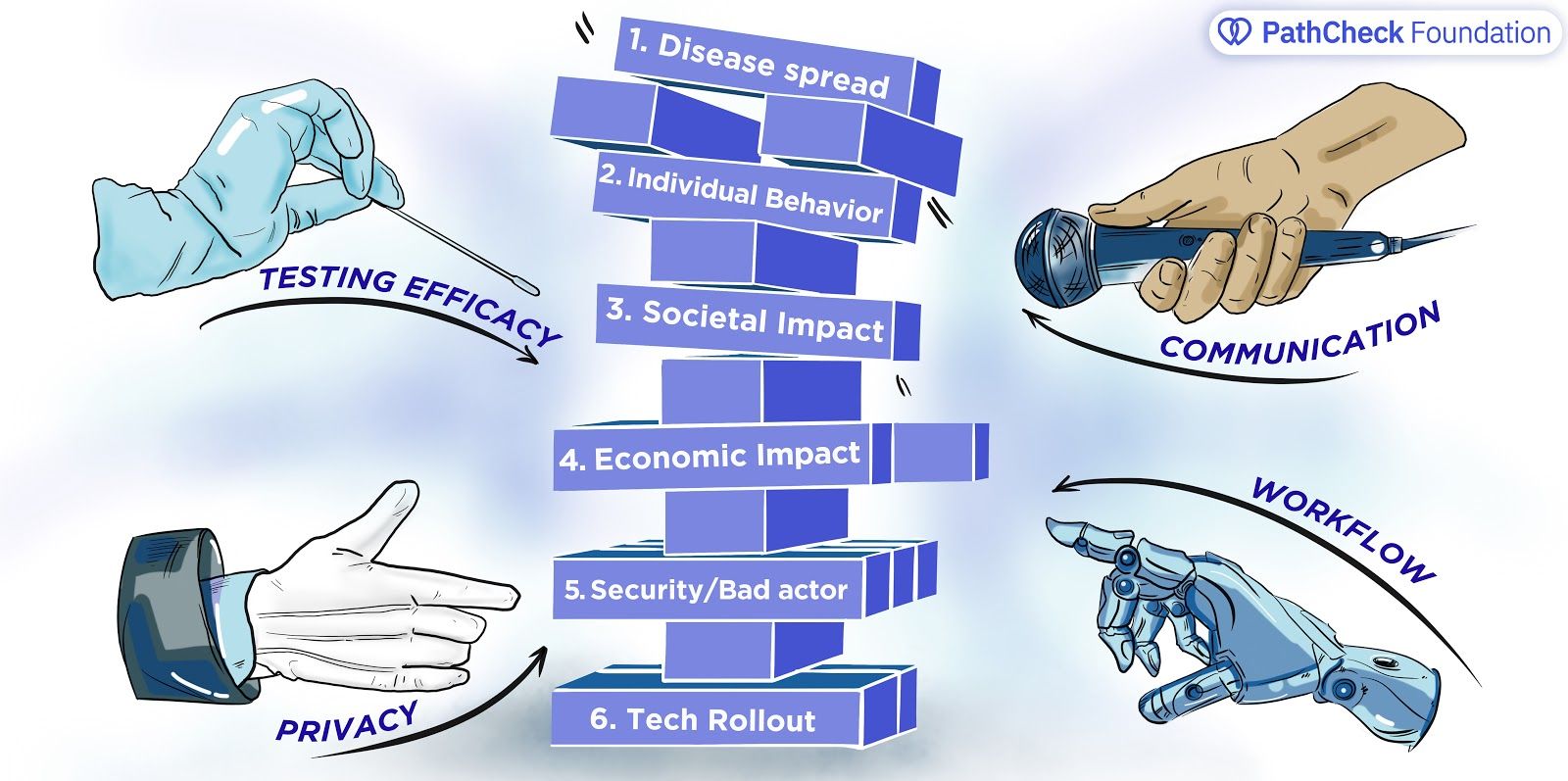}
    \centering
    \caption{Testing program utility is weakened as consequences of the four challenges remain unaddressed similar to how a Jenga tower is weakened by players removing blocks.  When enough consequences remain unaddressed, the testing program is compromised to the point of ineffectiveness. In the same way, once enough blocks are removed, the tower collapses.}
    \label{fig:jenga}
\end{figure}

\subsection{Privacy}
In this section, we consider the role of privacy issues and concerns in COVID-19 testing. Governments and health authorities collect massive amounts of personal data from people undergoing COVID19 testing including protected health information.  Testing, itself, generates a highly sensitive piece of data- the test’s result.  During the rush to enact testing programs, compromises to personal privacy have been made. While in some cases measures are taken to protect private information, gaps in privacy protection remain for many test users.  Left unaddressed these issues not only jeopardize the success of testing programs, but also creates unintended ethical and societal consequences, such as discrimination, and weaken social justice structures.

\subsubsection{Issues}
\textbf{Most test programs require the user to share more sensitive information while scheduling testing and receiving results than is strictly necessary to perform the test.} Users are asked or required to provide private information including their name, email, phone number, ethnicity, country of birth, social security number (in the U.S.), residential address and recent travel history. \cite{Walgreens} \cite{TuftsMedicalCenter}

\textbf{Test programs often lack transparency about the use and storage of the user’s data.}  Sensitive data may be retained by private testing companies, government-run public health testing programs, and employers through employer-sponsored test programs after the test is completed.  

\textbf{Many programs do not offer users an option to opt-out of sharing sensitive information and consenting practices are often inadequate.}  Good privacy practices dictate users should provide consent for data utilization.  Consent is often compromised by difficult, technical language and a lack of real choice.  In most cases, the user must provide their sensitive information in order to access the test.  

\textbf{Limited data security risks exposure of the identity of people with COVID-19.}  Data collected on paper forms often sits at test campsites.  Some programs convey results to users through websites that require no more than name and birthdate to look up the works. \cite{Ford}

\textbf{Large data sets create targets for bad actors.}  Health institutions and protected health information are frequent targets for cyber attacks and hacking.  Long term storage of test user data increases the likelihood the data will fall into the hands of bad actors. \cite{Yulisman}

\subsubsection{Addressing Privacy Challenges}
Countries around the world have taken heterogeneous approaches to the privacy challenges created by the response to the COVID-19.  Both the legal landscape and practical application vary.  In the United States the COVID-19 Data Protection Act of 2020 has been introduced in Congress.  If enacted, the act would supplement HIPAA protections for user’s protected health information and restrict “covered entities from collecting, processing, or transferring an individual's personal information for contact tracing with respect to COVID-19” without affirmative consent. \cite{Congress} 

Deployment of certain strategies has been delayed or canceled because privacy concerns could not be mitigated.  In the Canadian province of Quebec for instance, a very promising non-binary exposure notification mobile application was developed by a multidisciplinary team led by the Quebec Institute of Artificial Intelligence (MILA). The mobile solution made use of state-of-the-art machine learning algorithms to produce an unparalleled level of personalization in the recommendations displayed to the user. In order to do this, some personal data had to be collected on a regular basis for the model to perform and although high standards were used to encrypt, anonymize and store such data, this remained a subject of concern. A public largely mediatized debate ensued and the government decided not to support the deployment.

\subsubsection{Consequences of Privacy Challenges}
\textbf{Disease}: People’s perception on data privacy protocols plays an important role in the social adoption and acceptance of testing which has a direct impact on prevention of disease and mitigation of disease spread. People will be reluctant to participate in testing programs that are unnecessarily intrusive in terms of their data collection or that give the impression of being intrusive. Some individuals will not easily have all required information/documents (insurance card, driving permit, etc.), and marginalized populations may not participate if they have to divulge information due to mistrust and fear of stigma and discrimination.  Populations with high levels of data rights awareness will be similarly reluctant to participate. 

Regardless of the source, any hesitation to undergo testing will increase the spread of COVID-19 by delaying identification of cases and increasing the number of unidentified cases in the community.  

\textbf{Individual Behavior}:
Data privacy concerns limit our aggregate access to information and unintentionally facilitate  risky individual behavior. In the absence of an efficient testing program, individuals are constrained to making uninformed decisions about social distancing and quarantine and more likely to take risks.  Every-day decisions involve somewhat conscious evaluation of the tradeoff between risk and our individual utility function. Having less access to information does not prevent people from making decisions, they keep making those decisions, but the decisions will be uninformed, thus not necessarily optimal. 

\textbf{Social Impact}:
Mishandling of test results and data collected on testing sites or failure to preserve anonymity during the process alienates individuals and communities and leads to an exacerbation of systemic discimination and stigma. Further stigmatization and even discrimination against systemically margnialized individuals in public and work spaces amplifies exsisting inequalities.  

In the workplace and other public spaces, testing programs in combination with digital technology can have a negative impact to the extent that they open the door to abuse and misuses in direct confrontation with work law as well as basic work and employment rights. Private information leaks in the workplace cause unnecessary distress and jeopardize income and employment. Long lasting implications of these unintended consequences will extend beyond the pandemic and into the future. See report from the International Technology Law Association as a reference:   https://www.itechlaw.org/technology-governance-time-crisis

\textbf{Economic Impact}:
The economic impact has been drastic since when the world was hit by COVID-19, hard choices have been made and even harder sacrifices are yet to come. Economies have been hurt worldwide and there are few examples of jurisdictions or communities in the world that have successfully curb the spread of the disease. There is a sense of urgency to deploy as quickly and as many mitigation actions as possible. Testing programs is one such action that unequivocally can reduce spread rates and above all, that can serve as the cornerstone of comprehensive strategies allowing the gradual and safe opening of our economies in a world where vaccines are yet in the future or slowly been rolled out. It is at this moment, when the resilience of our economies have been put to the test for longer than anyone imagined, that decision makers might rush to deploy mitigation programs. Proper protocols for data collection, use and handling must nonetheless be put in place to avoid unintended consequences. Decisions made for the sake of the economy might have the exact opposite effect. Testing programs must not represent an economic burden for heavily affected sectors such as SME’s (Travel, food and entertainment businesses for instance). In the presence of a strategy that inadequately addresses the principle proportionality when implementing measures, small players would not be able to carry recurrent costs associated with such opening strategies that rely on routine and frequent testing. A testing program must then address privacy challenges in a cost effective way as not to inflict further damage to heavily hit economy sectors. 

\textbf{Security and Bad Actors}:
Testing sites and their data collection processes can become weak links in a security context. Personally identifiable information is now being collected during testing and it often sits on web servers or in handwritten forms at testing campsites. Due to COVID-19, phishing operations have become more frequent and there is a significant shift to target major businesses, government and critical infrastructure.  Large sets of sensitive data are attractive to hackers, threatening the digital safety of the individual.  Insufficiently restricted access to private data can lead to misuse.  Physical paper forms are subject to dumpster-diving and information can be easily stolen, putting details of individuals in the hands of bad actors. Eventually, as testing results begin to be required to have access to services or public spaces, forgery can also become an issue.    

\textbf{Technology Rollout}:
As technological solutions are developed to assist with the rollout of testing programs, they must take into consideration privacy preserving mechanisms as well as sound security measures to protect any data collected. Attempts to address these challenges will definitely increase the complexity of digital tools for COVID-19 testing but are nonetheless crucial. When technology is not perceived as being privacy preserving, societies might collectively choose not to use it. One recent example being the AI-based contact tracing application developed in Quebec by MILA. Although it had built-in multiple anonymizing protocols and safety features against bad actors, it was judged to be too high a risk.  Increasing complexity increases cost and lengthens timelines.  OS-dependency limits access to tools.  Data silos, created to protect privacy, limit the utility of data for public health and restrict user record linkage. All of these represent particular challenges when taking into consideration privacy concerns during the conception of a technological rollout.  

\subsection{Test Efficacy}
Medical tests are never 100\% accurate and thus false negatives as well as false positive rates determine the efficacy of the test. Inherently, COVID-19 testing programs face efficacy shortcomings that must be taken into consideration. Below we discuss main issues related to the efficacy of the testing program landscape. The extent to which each of the following issues complicates testing programs varies by test method used and time of testing from the appearance of COVID-19 symptoms.  However, the following issues can be frequently found in programs using currently available methods of testing.  

\subsubsection{Issues}

The testing landscape is a complex one as multiple tests, each one exhibiting its own advantages and shortcomings, co-exist. Antigen tests are fast and cheaper to use but are not as sensitive as the RT-PCR. While they can be used in vast numbers, concerns arise when the test will miss COVID-19 positive people who will continue the chain of infection. This highlights the need to develop testing strategies focused on reducing variation and improving efficacy. For instance, Slovakia recently instituted a two test approach to overcome the challenge of low antigen test sensitivity.  Slovakia tested 3.6 million people out of a total population of 5.5 million people using a swab-based antigen test that delivers results in less than 30 minutes.  Because the rapid test has a lower sensitivity than RT-PCR and likely misses a number of COVID-19 cases, Slovakia retested more than 2 million people within a week.  The country hopes the double testing plan will limit the number of false negative cases and decrease the spread of disease.  Preliminary results appear promising, infection rates in one community dropped from 1.47\% to 0.62\% over the course of a week.  Double testing may be one strategy to overcome the lower sensitivity of rapid tests while obtaining results quickly and less expensively\cite{programmed_review}.

\textbf{Both false negative results and false positive results occur.} Different types of tests\cite{fda_testing} have different levels of sensitivity and specificity\cite{testing_effectiveness} complicating the interpretation of a test result\cite{rethinking_covid}.  

\textbf{Test users receive results as a binary, yes or no, result.} Currently the tests are reported as a binary outcome while the underlying methods operate on a different level of thresholding\cite{duration_infection} to predict the outcome. The cycle threshold score for the RT-PCR test requires a certain extent of interpretation\cite{interpreting_covid_tests} that might not be suitable for the general public, nevertheless, giving a binary outcome can create a sense of panic or irresponsibility among a fraction of them.

\textbf{Long turnaround times for test results decrease COVID-19 test utility.}  Average test turnaround times for COVID-19 clinical tests were 2.7 days in September 2020, having fallen from 4.0 days in April 2020.  African Americans and Latinos had disproportionately higher average wait times; an indication of how minority groups may be medically underserved\cite{survey}. 

\textbf{The predictive value of the test changes over time.}  The likelihood of a false negative result varies based on the timing of the test relative to the person’s exposure or onset of symptoms.  A recent analysis by Kucirka et al suggested the likelihood of a false negative result from an RT-PCR test on the day of exposure is 100\%; the likelihood of a false negative decreases to 20\% by day 8 post exposure\cite{acp_article}.  

\textbf{Record linkage is incomplete, limiting public health analysis.}  During the course of the pandemic, some people will undergo testing more than once.  Linking their test records provides critical information about the spread of disease within a community.  However, anonymization, lack of common identifiers, and data silos frequently prevent this\cite{review_programme}. 

\subsubsection{How Issues with Test Efficacy are Impacting our World}

Slovakia recently instituted a two test approach to overcome the challenge of low test sensitivity.  Slovakia tested 3.6 million people out of a total population of 5.5 million people using a swab-based antigen test that delivers results in less than 30 minutes.  Because the rapid test has a lower sensitivity than RT-PCR and likely misses a number of COVID-19 cases, Slovakia retested more than 2 million people within a week.  The country hopes the double testing plan will limit the number of false negative cases and decrease the spread of disease.  Preliminary results appear promising, infection rates in one community dropped from 1.47\% to 0.62\% over the course of a week.  Double testing may be one strategy to overcome the lower sensitivity of rapid tests while obtaining results quickly and less expensively.\cite{programmed_review}
    
\subsubsection{Consequences of Test Efficacy Issues}

\textbf{Disease}: False negatives, misunderstandings about the timeframe in which testing should be completed, and other challenges with test efficacy work together to decrease the utility of testing.  If unaddressed, these challenges result in greater spread of disease as infected individuals go unidentified and fail to isolate.
 
\textbf{Individual Behavior}: One desired effect of a testing program is that of inducing a change in our collective behavior. Our daily decisions should ideally be influenced by the information that individuals gain through the use of testing programs. If test efficacy is an issue, then the mitigation strategy is not influencing individual choices the way it should be. Challenges with test efficacy decrease the quality of information individuals receive from a COVID-19 test leading to uninformed or misinformed decisions about social distancing and quarantine.  Individuals receiving a false positive result experience stress and possibly financial loss if their quarantine impacts their employment.  Confusion over the utility of tests and the accuracy of results creates mistrust and hinders adoption of testing programs.    

\textbf{Social Impact}: Testing efficacy is a key element in the ethical and societal impacts of mitigation strategies. The Trade-off between benefits and risks is central to ethical considerations when rolling out a given strategy. As we have seen, there are challenges and unintended consequences associated with a testing program rollout. By their very nature, these uncertainties introduce risks that can potentially have negative impacts in the fabric of our societies. In the absence of a clear perspective on the accuracy of results, it does not seem ethically viable to expose our society to the associated risks. Confusion stemming from poorly addressed challenges of test efficacy decreases compliance with public health recommendations and contributes to the spread of rumors and falsehoods.  Inaccurate results put some test users at increased risk for social stigma.  Failure to account for the challenges of test efficacy decreases the accuracy and utility of public health data and analytical tools, potentially resulting in poor policy choices or poor program design. These risks should not be underestimated as they are linked to important ethical questions that can also be decisive in the success of a testing strategy. If accuracy and efficacy of the testing program cannot be guaranteed, societies may choose the longer term pursues of social justice over short term inadequate solutions or strategies hence resulting in a miss opportunity to curb the spread. The choice of the government of Quebec not to roll out a AI-based non-binary exposure notification strategy was also based on a lack of overwhelming quantifiable evidence of its efficacy that would justify the risks of negative social impact of a roll out. 

\textbf{Economic Impact}: Understanding of test accuracy, pitfalls, false positive rates, false negative rates, and result interpretation are key to the design of economy recovery plans.  Steps to alleviate the impact of test efficacy challenges, such as repeat testing, add cost to testing programs.  As mentioned before, testing programs with high costs could potentially bring more damage to the SME sector. Long turnaround times and false positives delay the reopening of businesses and the return to work of exposed employees.  Such delays are particularly devastating to workers without sufficient paid sick leave and low income workers.   

\textbf{Security and Bad Actors}: As testing becomes more widespread and embedded into our daily lives as a requirement to integrate economic and social activities, more information will be shared and new protocols will be put in place. In this context, people run a higher risk of exploitation and misuse of the information. Testing programs that do not introduce mechanisms via double testing will more likely create situations where bad actors find opportunities to exploit information for instance by conducting forgery activities around testing results certificates. 

\textbf{Technology Rollout}: Using technology solutions to assist with the deployment of testing programs will face the challenge of integrating strategies designed to improve efficacy such as double testing. Challenges in test efficacy complicate analysis and use of data by public health and, if not accounted for, decrease the quality of data analytics.  Attempting to address these challenges may require greater data sharing by the user, potentially impacting privacy.  Accounting for details, such as the day of testing relative to the day of exposure, testing site conditions, and methodology, all increase the complexity, and, therefore, cost and timeline, of the development of a digital tool. Such complexities also increase the likelihood of OS dependence which can leave out those without consistent and individual access to smartphones.  

\subsection{Test Supply and Workflow}
Countries and states are persistently facing challenges related to the supply of testing kits and management of testing capacities. 

\subsubsection{Issues}

\textbf{Inconsistent testing policies create confusion and inefficiencies. } Recommendations for when to undergo testing vary between jurisdictions [ref].  Some test program policies address challenges with test efficacy, such as optimal timing post exposure for undergoing testing or variations in test specificity and sensitivity between methods while others do not.  This patchwork of policies persists in spite of guidelines from leading organizations.

\textbf{Limited testing capacity stymies test program deployment. } Shortages in testing capacity have resulted from shortages of laboratory equipment, testing swabs, and other test supplies, shortages of trained personnel to perform the tests, limited numbers of testing sites, and data management issues. Mismatch between test site capacity and the number of users directed to the site further contributes to capacity issues.  In May 2020, the number of tests performed daily utilized only 70\% of the global testing capacity.  

\textbf{Cost of testing, or uncertainty of the cost, reduces uptake of testing. } Lower costs increase testing uptake.  In the US, the Families First Coronavirus Response Act requires insurance to cover COVID-19 tests with no costs to the user; coverage is not mandated for the uninsured, though some states are offering free testing to this population.  Uncertainty about coverage of related services and a fear of unexpected bills remain barriers to testing. Cost concerns also hinder deployment of new testing programs. Whereas, some schools have been able to implement large testing programs, others are struggling to fund programs.

\textbf{Individuals struggle to access testing sites. } Programs directed people to testing locations 250 miles from their homes.  Others sent family members to different testing locations as a result of insufficient testing capacities and supply chain disruptions. These difficulties compound any anxiety the user may have about testing.

\textbf{Data management and IT problems plague some testing programs. } Testing programs generate large data sets.  Officials struggle to manage test results, and link the data across jurisdictions.  16,000 positive test results went unreported when the Excel sheet used to track cases grew too large to successfully transfer to the central computer system in the UK.   Such errors contribute to the mismatch between testing needs and testing supplies in some regions. Human errors further contribute to the chaos.  Test results were sent to the wrong person following incorrect entry of the person’s phone number into the reporting system.

\subsubsection{How some countries are deploying (background)/ references to real world projects}

Institutions, countries and companies are deploying various solutions to tackle supply and capacity management challenges. The Vulnerable Population Dashboard developed by McKinsey can help officials understand regional populations and allocate supplies in line with the demand by identifying geographic areas with relatively higher concentrations of people who may be vulnerable to COVID-19 and COVID-19 related public health measures based on their age, physical or behavioral health conditions, access to healthcare services, and social factors.[ref] US Digital Response Team’s online dashboard, created in partnership with the Pennsylvania Department of Health to track the availability of hospital beds and ventilators on a county-by-county basis, could serve the purpose of tracing the supplies of vaccines and testing kits. In Massachusetts, a mobile testing unit, financed by the state, is made available to private and public schools upon request in order to perform rapid testing of a group of students and/or teachers and staff once transmission has been detected and identified under certain, specified conditions. Free testing will be performed on asymptomatic students or staff that might have been in close contact with a person tested positive for the virus.[ref]
  
\textbf{The EU released guidance in an effort to unify the testing approach across the European Union [ref].} The document focuses on highlighting action points for countries when adapting their national testing strategy for different stages of the COVID-19 pandemic response, covering such issues as: types of tests, testing of symptomatic and asymptomatic cases, testing capacities, testing turn-around-time, testing in specific settings: schools, travelling, contact tracing and mobile applications.

\subsection{Consequences of Testing Management Issues}

\textbf{Disease}: Spread of the disease is a consequence inherently linked to testing workflow and logistics challenges. The issues with access to testing and capacity management problems negatively affect the ability of health authorities to efficiently and effectively perform testing and contain pandemic. Insufficient testing and a decrease in the uptake of testing by the population as a result of mistrust caused by logistics issues increase the number of undetected or late detected cases in the community causing further spread of the pandemic.

\textbf{Individual Behavior}: Challenges with the logistics and workflow of testing programs (inconsistent policies, limited access to testing sites, cost, etc) all combine to make it more difficult for people to access testing.  When a person is unable to access testing, he/she is then making uninformed decisions in his/her daily lives potentially decreasing adherence to isolation guidelines despite most people’s intent to follow guidelines.  Challenges with logistics also increase complexity of testing programs and confusion for potential users leading to a decrease in trust.

\textbf{Social Impact}: Matching supply with a demand for testing, in particular in the hot spots, where most vulnerable populations live, is crucial to address the already persistent racial and ethnic inequality in health care delivery. The failure to do so, contributes to even greater social disparity. The inability to test, diagnose and provide necessary health care to vulnerable populations, would foster negative behaviours and adherence to test, trace and isolate, and in the medium turn will lead to negative economic impact for those populations. 

\textbf{Economic Impact}: Inefficient supply chains, inaccurate testing, need for retesting, varied and over-the-top cost of test kits, and lack of coordinated reporting of supply increase the cost of testing operations. This indirectly affects economic recovery initiatives. Inefficiently operated testing programs represent an inefficient allocation of resources. An inefficient testing program operation slows down any attempt to deploy economic recovery measures.  Delayed access to testing limits the ability of individuals exposed to COVID-19 to return to work or school in places where a test result is required for return thus impacting economic recovery. 

\textbf{Security/Bad Actors}: Inefficiencies and mismanagement of test data increase the risk for security breaches. Such issues create opportunities for bad actors to exploit. By multiplying the number of weak security points, inefficient workflows represent a threat not only to individuals but also at a national level where systematic attacks on a country’s COVID response can weaken that country’s path to recovery.

\textbf{Tech Rollout}: Inefficiency and complexity in test program logistics increase the complexity of technological solutions attempting to manage test programs.  Cost increases along with complexity. The software could actually help to establish a better and concrete structure to address the issues of test logistics such as the initial registration to get tested, alerts about the new test methodologies and most importantly the process can be carried out smoothly but giving specific time slots to the individuals for them to get tested, this would not only help fastrack the process but also ensure the safety of the patients who are visiting the test sites. But none of this is possible if underlying workflow logistic and workflow challenges are not addressed beforehand and the solutions integrated into the development of the technology. 

\subsection{Communication}

Success of mitigation strategies in Today’s democratic societies will depend heavily on social adoption and acceptability of the different components that constitute it. In the absence of a real possibility to enforce a given strategy as compulsory, the main vehicle to success is to effectively influence individual choices to the point that collective behavior is modified and steered towards safety patterns. Communicating not only the objectives of a given strategy but also how it operates as well as its blind spots is key to increasing acceptability and adoption. Communication must be comprehensive, transparent, inclusive and effective in order to create the trust that is required at the individual level for it to translate into a significant behavior shift at the collective level. Communication is also key for the coordination and logistics required to effectively roll out a strategy at the scale required to significantly reduce the spread. For instance, communicating interpretation of test results in ways that effectively empower individuals with enough information to make safe decisions as they move about in their daily lives during the long months that are required to safely open our societies again. Communication efforts must also look to unify different channels and reduce misinformation. Information outputted through media channels about the COVID-19 pandemic has been  at its peak for the past couple of months with no sign of receding. Alongside valuable scientifically or otherwise vetted information, one finds non-scientific  speculation, heavily politicized claims, all sorts of innocent rumors all the way to outlandish conspiracy theories. Infodemics, which can be described as an abundance of information, is very prevalent with respect to the COVID-19 crisis and this is a background chatter out of which a clear message must rise above in order to be heard and be efficient. This overabundance of information has a non negligeable psychological impact  and makes people skeptical as to what source to believe and who to believe. It is usually hard to distinguish the facts from the rumors and this excess of information can be very taxing to an individual as well. This in turn can affect an individual’s mental health, adding up to his/her stress and causing immense anxiety and pressure leading to a nonchalant attitude and behaviour at the individual level. In addition to individual perplexity, communities become vulnerable to the risk of polarizing opinions and hinder their engagement towards holistic improvement efforts.   The COVID-19 pandemic has been exceptionally stressful at many levels representing an inedit situation bringing a heavy toll on everyone as well as consequences never seen before and that hinder future mitigation strategies.

\subsubsection{Issues}
\textbf{Precise and reliable communication from officials to the public. }The ongoing worldwide COVID-19 crisis is posing tremendous challenges for government officials on many fronts as they have had to coordinate large scale responses to contain the immediate health threat posed by the pandemic. The many dimensions of these efforts makes it of uttermost importance to have a unified, reliable and precise communication from government officials to the public. This communication is key in dealing with the pandemic. Additionally, the public health messaging is not just PSAs but employing other media outlets to debunk fake news and simplify scientific developments for adoption of nanopharmaceutical interventions. Decisions and policy must be clearly communicated in a transparent and understandable way so as to not create mistrust hence maintaining the capability of officials to influence behavior. .Furthermore, health equity challenges limit uniform distribution and reception of key messages. 

\textbf{Changing messaging over time (science) is difficult to communicate well.} The very nature of the scientific process has self-correcting mechanisms that, using the tools and methods from science, improve our understanding of the world in an evolving everchanging way. In any scientific endeavour, complete certainty is never achieved and thus in a never-seen-before situation, where our understanding of the different layers of the problem evolves with the self-correcting mechanisms of science, we are learning and adapting on the fly and so are our policies, recommendations and strategies. Given the amount of uncertainty, the advancements in technology, the immense number of people contributing towards the cause, there are numerous discoveries and updates made available every day which. This will likely lead to messages and recommendations being modified as our understanding improves. What holds valid today, might be better understood in the future and not have the same significance tomorrow. These continuously varying messages from government officials can cause disbelief or misunderstanding amongst the public and can result in masses discrediting government policies and can cause mistrust that hinders the ability of officials to influence behaviour. 

\textbf{Scientific literacy is very crucial on the policymakers, government official's end as well on the part of the masses. }Having scientifically literate team members and/or spokespersons in the communication team is required in order to design a coherent message that the public can understand and rally behind the implications of various policies and how the action plan is implemented. At the same time, scientific literacy can assist government officials in making well-informed decisions. Scientific literacy can help in bridging the communication gap between the various entities involved in the system and the public in general and the lack of it can be detrimental to everyone. [link]

\textbf{A coherent and unified message must be voiced by all levels of government and by different officials within their agencies and various organizations.} In complex situations like the current sanitary crisis we have witnessed mixed messages and statements regarding the severity of the covid-19 crisis and later of recommendation and the mitigation strategy. It is essential that all the government agencies and systems worldwide voice their opinions in a coordinated effort and as part of a comprehensive plan. Consistent communication from the government officials backed by factual information can be beneficial and help the public understand better the magnitude of the situation better thus empowering them to make the right choices. [link]

\textbf{Politicization of the sanitary crisis brings about supplementary difficulties that endanger the success of any mitigation response hence unnecessarily putting people’s health at risk.} The primary objective of a mitigation response must be that of saving lives, influencing individual behavior in ways that effectively reduce the spread and ultimately creating the conditions so that our societies can safely reopen. Efforts to hijack the message and actions of a mitigation response for political gain not only is unethical, since it prioritizes other objectives over human well-being, but it will also result in polarizing the public thus hindering the ability of the message to influence collective behavior at scales that can have a significant effect in reducing risk. More than half a year into the crisis, we know all too well the negative impact in scenarios around the globe where, because of political motives, the risk of the pandemic is downplayed or the opinion of healthcare experts is disregarded or put aside from the decision making process.  Needless to say the message must be devoid of political motives or any other that does not prioritizes reducing the spread and its negative impact on our society, economy and communities. 

\textbf{Test results are non-binary: Interpretation of results needs to be non-binary as well.} COVID-19 test results are often perceived in binary terms, either a patient is infected or not. It needs to be taken into account that there are many more layers of complexity to the test results than just a positive or negative result. The notion of non-binary results must be communicated effectively as to create awareness about the possibility and mening of false positives and false negatives. Indeed, in order to endow the public with the right information, thus empowering them to make the right decisions, it is essential that test results are understood in terms of likelihoods rather than certainty. There is also the time dimension of the test and the individual timeline from the day of contagion of a given individual. Viral loads evolve in a statistically predictive pattern and so the efficiency of a given test in detecting the presence of the virus. Depending on the particular context and timeline of an individual transmission, there are time windows for which a given test results are less accurate or relevant. Not all tests are equal nor constant in time in terms of accuracy, false positive or false negative. These creates a challenge when communicating this message as part of a mitigation strategy.      

\textbf{Social media is influencing the information that individuals consume greatly.} There have been various researches carried out claiming huge percentages of the population believed in misinformation after solely being influenced by social media posts and articles. The contribution by social media toward misinformation and rumors are not regulated. Opinion pieces, written by individuals are perceived as facts. Awareness regarding the implications of misinformation and spreading the same is very crucial. 

\subsubsection{Examples of how the world is dealing with Communication Issues}
 
\textbf{Efforts are being made worldwide to handle issues related to communication in regards to the COVID-19 pandemic. }The UN set-up a COVID-19 Communications Response Initiative to coordinate the efforts of the different Agencies, establish partnerships to debunk fake news and rumors with scientific and solution-based information. “Fear, uncertainty, and the proliferation of fake news have the potential to weaken the national and global response to the virus, bolster nativist narratives and provide opportunities for those who may seek to exploit this moment to deepen social divisions,” said Melissa Fleming, Under-Secretary-General for Global Communications “All this threatens to undermine the international cooperation urgently needed to deal with the impacts of this crisis.” \cite{un_infodemic} WHO created the Information Network for Epidemics (EPI-WIN) merging the social media and technical teams to monitor the information flow and react quickly to misinformation with fact-based material. \cite{EPI_WIN} It partnered with Viber, Facebook, and WhatsApp for a multilingual messaging system to be able to directly reach out to 2 billion people. It offered key resources and access to events to journalists and denounced restrictive measures and attacks from certain governments towards the media. It mobilized governments and civil society organizations to diffuse locally official information from the WHO. In order to help people sail through this large quantity of information and help to flatten the COVID-19 infodemic curve, WHO initiated a Mythbuster program. It aims at clearing up all the most common myths and misconceptions around COVID-19 by listing by providing tips on how to recognize false information. \cite{mythbusters} "This ‘infodemic’ is proliferating and potentially deadly. We cannot cede our virtual spaces to those who traffic in lies, fear, and hate." says Antonio Gutierrez, Secretary-General of the United Nations. The UN launched the "Verified Initiative" to curb misinformation around COVID-19. It is calling on people worldwide to sign up and act as "information volunteers" to distribute verified content in their communities. The UN Department for Global Communications will send to the volunteers a daily feed of verified and easy to share information material. The ultimate goal is to contrast misinformation or fill information voids. \cite{good_comm_saves_lives}  Combating rumors about COVID-19 in areas where digital resources are limited, such as refugee camps, does not involve advanced artificial intelligence. The International Organization for Migration (IOM) is training refugees in Cox's Bazar in Bangladesh to use bicycles and loudspeakers to deliver precise news and instructions. Message material goes from scientific facts on COVID-19 to general information on mental health and psychosocial assistance. The messages are kept on USB drives so that data can quickly be modified to different situations.  The initiative also aims to reach the refugee population not covered by the UN Agency's earlier project called COVID Info Line that allowed users to record questions, comments, and share concerns \cite{battle_misinfo} \cite{COVID_info_line}. Patwa et. al. 2020 \cite{patwa2020fighting} released a dataset of COVID-19 related fake news on social media. 

\subsubsection{Consequences of Communication Issues}

\textbf{Disease}: Errors in communication reduce the efficacy of COVID-19 testing programs and generate mistrust.  Communication failures create barriers to testing and reduce community compliance with public health directives leading to increased spread of COVID19.  COVID-19 is an emerging disease and there is more to understand about the its’ progression, transmission and severity. Timely and consistent public health communications encourages institutional preparedness and coordinated efforts for availability and accessibility of testing .

\textbf{Individual behavior}: Miscommunication results in sub-optimal actions.  Individuals bombarded with mixed messages feel frustrated and uncertain.  Unsure of whether or not they should undergo testing or unable to quickly determine how to be tested, many will choose not to pursue testing.  The risk individuals will not undertake testing when indicated is increased by fears of stigma and discrimination, either by their community or their employer. Communication errors also result in misunderstanding of test results by the person undergoing the test.  As a result, an individual with a significant risk to be infected may not properly isolate, increasing the likelihood they infect another member of their community.  The impact of miscommunication on the mental health of individuals must also be considered.  Feelings of fear or panic, stress, and frustration all negatively impact mental health and increase the burden on those with mental illness. The stigma associated with social isolation is a deterrent for testing. Lack of targeted messaging according to the age groups or vulnerability levels creates a complex pool of information, difficult to decipher and hence adversely affecting the adoption of testing, when needed.    

\textbf{Societal}: Miscommunication spreads information inconsistent with the views of public health authorities and results in a lack of compliance with critical recommendations and mandates. It decreases buy-in from community leaders; both official and unofficial leaders are hesitant to endorse or promote ideas when the risk for error is high. Confusion and miscommunication among officials can be incorporated into policy decisions, magnifying the impact of the communication error.  The chaos caused by miscommunication is fertile ground for the development of rumors.  Across the globe, we have seen the serious, fatal, impact of untrue rumors spread rapidly through a community or across social media platforms. Discrimination builds on the foundation created by rumors. Such an extensive and serious concern, that the United Nations has issued guidance on addressing and countering COVID-19 related Hate Speech as well..  Lack of focus on interracial and culturally sensitive communcations disporporationaltly affects health outcomes and timely care utilization. As marginalized communities are underrepresented in healthcare, there is a baseline for low levels of trust and collective fear, making communication even more difficult. 

\textbf{Economic}: Miscommunication about testing threatens the economic health of a community. Employer misunderstanding of COVID-19 testing strategies, testing protocols, or test results leads to incorrect decision making about employees’ ability to remain at work. Employees who should be isolating may still be required to report to work due to miscommunication about the meaning of a negative test result risking the further spread of disease within the workplace. In other cases, an employee may be required to remain out of work due to a miscommunication about their risk for carrying COVID-19. For employees without access to paid leave, this may be financially devastating. The employer may also suffer if forced to close due to insufficient staff levels. If COVID-19 testing programs are seen as unhelpful, poorly managed, or ineffective due to communication errors, employers are less likely to implement these into their workplaces or be willing to release critical data about COVID-19 infections within their workplace to employers. Miscommunication about testing also impacts the community’s perception of the business, potentially resulting in community members avoiding the business despite the low risk of disease transmission at the site. In worst-case scenarios, businesses have been forced to close.     
The economic value of COVID-19 testing far surpasses the cost of the test. Hence, miscommunication hindering testing creates an economic burden on the health system.

\textbf{Security}: Miscommunication creates space for bad actors. Fraud is easier to perpetrate in a climate of uncertainty and frequent falsehoods. \cite{Samuels_and_Kelly} Particularly in the context of health verification, misunderstanding of test results, the necessity of testing, and the impact of testing encourage misuse, particularly if the health verification is tied to employment or access to desired services. Individuals are more likely to be targeted when misconceptions about the disease are prevalent, possibly endangering them and their families. Miscommunication also increases skepticism and concerns about privacy leading some individuals to decline to participate in testing. Similarly, contact tracing approaches are also prone to similar security breaches barring willingness to participate in testing.   

\textbf{Tech Rollout}: The uptake of technological tools for COVID-19 testing is complicated by information fatigue and mistrust resulting from miscommunication.  Individuals unable to decipher which tools are secure and accurate are likely to scorn all available tools.  The utility of health verification tools, in particular, will be tied to community trust in the tools.  Miscommunication threatens to eliminate the impact such tools might have on the control of the pandemic. This takes away the individual ownership of health status prompting susceptibility to rumours and miscommunications. 

\subsection{Increased Impact of Challenges on Marginalized Populations}

Testing programs which fail to adequately address the challenges described in the preceding sections often exacerbate existing inequities as marginalized populations face an increased magnitude of the consequences than individuals with greater social, economic, and political status.  In the United States, communities of color are experiencing unequal incidences of COVID-19 infections and stark differences in mortality rates compared to White Americans \cite{CDCracial}. Indigenous, Black, and Latinx Americans were at least 2.7 times more likely to die from COVID-19 in 2020 \cite{apm}.  Particularly fraught for marginalized populations are data collection and privacy violations as these groups are often at greater risk of government surveillance and abuse than their non-marginalized counterparts \cite{newamerica} and already primed to mistrust government or health institutions due to past abuse \cite{Scharff2010}.   The remaining challenges may also disproportionately impact marginalized populations.  While Black and Hispanic/Latinx Americans are more likely to attempt to get ­­­­­­tested for COVID, they are less likely to be tested, due, in part, to test supply and workflow challenges \cite{mckinsey}, as well as persistent racial and ethnic disparities in healthcare delivery \cite{natresearchcouncil} and structural racism \cite{structuralracism}.  Each challenge must be addressed in a manner which accounts for the needs of vulnerable individuals and marginalized groups.  

These unequal negative effects are now very present in public discussions.  Some societies are making the choice not to roll out strategies because they deem these strategies pose too high a risk to the fabric of our democracies and to our collective pursuit of social justice. The province of Quebec, Canada, decided not to roll out a mobile application that made use of artificial intelligence and required the collection of a modest amount of private data. The current context and the discussions around the role of technology in mitigating the pandemic initiated an entire debate around data privacy and the role of technology in our daily lives and how regulation should be enforced in order to bring back to individuals the control of their digital data footprint into the future and beyond the current emergency.  These risks should not be underestimated as they are linked to important ethical questions that can also be decisive in the success of a testing strategy. In the absence of adequate protocols that preserve data privacy and anonymity, societies choose the longer term pursues of social justice over short term inadequate solutions or strategies.

%% file: content/discussion.tex
\section{Discussion and Conclusion}
Test, trace, isolate, vaccinate are key strategies in taming the pandemic. Our earlier work on ‘Apps Gone Rogue’, we considered unintended consequences of poorly designed exposure notification apps. In this paper, we have considered the unintended consequence of COVID-19 testing ecosystem due to challenges in privacy, efficacy and miscommunication. We believe the ecosystem is ripe for innovation. Given the tradeoffs between privacy, utility, efficacy and inclusivity, we realize there are no easy solutions. However, by considering the wider design space and their impact, our paper presents a useful blueprint for policy makers, testing labs and app developers in the current and future pandemic.